# Analytical scanning and transmission electron microscopy of laboratory impacts on Stardust aluminum foils: interpreting impact crater morphology and the composition of impact residues.

Anton T KEARSLEY[1*], Giles A GRAHAM[2,1], Mark J BURCHELL[3], Michael J COLE[3], Zu Rong DAI[2], Nicholas TESLICH[2], John P Bradley[2], Richard CHATER[4], Penelope A WOZNIAKIEWICZ[1,4], John SPRATT[1] and Gary JONES[1].

[1]Department of Mineralogy, Impact and Astromaterials Research Centre, Natural History Museum, Cromwell Road, South Kensington, London SW7 5BD, UK.
[2]Institute of Geophysics and Planetary Physics, Lawrence Livermore National Laboratory, Livermore, California CA 94551, USA.
[3]School of Physical Sciences, University of Kent, Canterbury, CT2 7NH, UK
[4]Imperial College of Science, Technology and Medicine, Exhibition Road, South Kensington, London SW7 2BP.
*Corresponding author. E-mail: antk@nhm.ac.uk

**Abstract-**The known encounter velocity (6.1kms$^{-1}$) and particle incidence angle (perpendicular) between the Stardust spacecraft and the dust emanating from the nucleus of comet Wild 2 fall within a range that allows simulation in laboratory light gas gun experiments designed to validate analytical methods for the interpretation of dust impacts on the aluminum foil components of the Stardust collector. Buckshot of a wide size, shape and density range of mineral, glass, polymer and metal grains, have been fired to impact perpendicularly upon samples of Stardust Al1100 foil, tightly wrapped onto aluminium alloy plate as an analogue of foil on the spacecraft collector. We have not yet been able to produce laboratory impacts by projectiles with weak and porous aggregate structure, as may occur in some cometary dust grains. In this report we present information on crater gross morphology and its dependence on particle size and density, the pre-existing major and trace element composition of the foil, geometrical issues for energy dispersive X-ray analysis of the impact residues in scanning electron microscopes, and the modification of dust chemical composition during creation of impact craters as revealed by analytical transmission electron microscopy. Together, these observations help to underpin the interpretation of size, density and composition for particles impacted upon the Stardust aluminum foils.

**Introduction**

The flight of the Stardust spacecraft through the coma of comet Wild 2 resulted in near perpendicular impact geometry for dust particles, at a relative velocity of 6.1 kms$^{-1}$ (Brownlee et al., 2003). In a suite of papers, Graham et al. (2006), Hoppe et al. (2006), Kearsley et al (2006), Leroux et al. (2006) and Stephan et al. (2006) have demonstrated that it should be possible to obtain important information about Wild 2 dust particle size and composition from craters on the aluminum (Al) foil exposed on the front surface of Stardust's particle collector. However, several questions remain to be answered.
Can a comparison between laboratory impact craters and the morphology of Stardust foil craters be used to help provide a measure of the overall structure and density of particles? Also, before residue analyses can be used to infer particle composition, it is important to document possible contamination by the substrate material. What are the analytical limitations imposed by the substrate? Energy dispersive X-ray analysis can provide a very rapid non-destructive impression of the major and minor element make-up of residue within craters of sub-micron and larger scale. Although X-ray data can be obtained from deep within craters, the crater shape and thin inclined residue sheet morphology generate difficulties with X-ray 'matrix corrections' for quantitative analysis. What is the best crater location for obtaining reliable quantitative analyses, and how should the sample be oriented? Is there substantial change in elemental composition from the impacting particle to the resulting residue? Can this be quantified, and is it dependent upon the size of crater and the volume of residue?
In this paper we seek to answer the above questions, to help alleviate uncertainties in the interpretation of the Stardust foil craters.

**Target and projectile materials, light gas gun shot conditions, imaging and analysis protocols**

Flight spares of Stardust Al1100 foil samples were utilised in all the laboratory impact experiments described below. Nominal foil thickness was 101 microns, and fifteen measurements of a vertical section through a 2cm wide foil sample showed an average of 101.3 microns thickness with a standard deviation of 2 microns. Foil targets were wrapped tightly onto a square of 1mm thick aluminium alloy plate to simulate the mounting on the Stardust collector (Tsou et al., 2003, Hörz et al., in press), then the plate was held by conductive adhesive putty upon an aluminium supporting sheet in the light gas gun (LGG) target chamber, perpendicular to the projectile trajectory.

Mineral projectile powders with a wide range of grain size (polydispersive) were selected and prepared from samples in the collections of the Natural History Museum (NHM), London. In this paper we report data from separate firings of olivine (specimen BM.1950,337; shot G220306#2 at 5.89 $kms^{-1}$), Pyrrhotite (BM.2005,M317; shot G210306#1 at 5.92 $kms^{-1}$), Diopside (BM.2005,M310; shot G210306#2 at 6.01 $kms^{-1}$) , Enstatite (BM.2005,M318 shot G270405#1 at 5.85 $kms^{-1}$), Bytownite feldspar (BM.2005,M312; shot G220306#1 at 6.01 $kms^{-1}$), and basalt glass (United States Geological Survey sample NKT-1G; shots G270505#1 at 6.10 $kms^{-1}$ and G270505#2 at 6.06 $kms^{-1}$).

In addition, projectiles of a range of known density and with very narrow size range (monodispersive samples) were purchased from:

1) 3M (bubble glass spheres, density 0.4 $gcm^{-3}$; nominal sizes 20-32, 53-63 and 63-75 microns diameter; shots G210406#1 at 6.00 $kms^{-1}$, G190406#1 at 5.94 $kms^{-1}$ and G240406#2 at 6.02 $kms^{-1}$ respectively)
2) Sigma Aldrich (poly-methylmethacrylate polymer spheres, density 1.19 $gcm^{-3}$; nominal sizes 30, 57 and 80 microns diameter; shots G090506#1 at 5.97 $kms^{-1}$, G040406#2 at 5.93 $kms^{-1}$, G080506#1 at 5.89 $kms^{-1}$ respectively)
3) Whitehouse Scientific (soda-lime glass spheres, density 2.4 $gcm^{-3}$; sizes and velocities as reported in Kearsley et al., 2006)
4) Salem Speciality Spheres (steel spheres, density 7.9 $gcm^{-3}$; nominal sizes 32-53 and 75-90 microns; shots G110506#1 at 5.87 $kms^{-1}$ and G 120506#1 at 5.96 $kms^{-1}$ respectively)

All the LGG shots with these projectiles were performed in the School of Physical Sciences at the University of Kent, Canterbury using the buckshot and velocity measurement techniques reported by Burchell et al., (1999).

Electron images of craters were generated using a JEOL 5900 LV scanning electron microscope (SEM), with X-ray spectra and maps acquired using an Oxford Instruments INCA Energy Dispersive Spectrometer (EDS), all at NHM. Quantitative X-ray microanalyses were performed at 20 kiloVolts (kV) and 2 nanoAmperes (nA). X-ray maps of Stardust and target foils were created by Wavelength Dispersive Spectroscopy (WDS) on a Cameca SX50 electron microprobe at NHM. Focused ion beam (FIB) preparation at Imperial College used a FEI FIB200 TEM instrument, with ultrathin sections extracted using an external manipulator. Sample preparation at Lawrence Livermore National Laboratory (LLNL) was performed on a FEI Nova 600 dual beam microscope comprising a Ga+ liquid metal source FIB and field emission gun SEM (FESEM). This dual beam microscope was fitted with an EDAX Genesis EDS and an Ascend Xtreme access nanomanipulator. Analytical transmission electron microscopy (AEM) was performed on a 200kV FEI Tecnai $G^2$ F20 XT Scanning Transmission Electron Microscope (STEM) fitted with EDAX EDS and FEI TIA spectral processing software at LLNL, California.

**Calibration of crater diameter as a function of both impacting particle diameter and density**

The existing calibration of crater diameter on 101 micron Al1100 foil as a function of impacting particle diameter at velocities close to 6 $kms^{-1}$ (Kearsley et al., 2006) was based upon sub-spherical projectiles made of soda lime glass, with a density of 2.4 $gcm^{-3}$, below the typical density for natural silicate minerals (e.g. olivine ca. 3.2 $gcm^{-3}$). To provide crater top lip diameter calibration for a wider range of impactor densities, especially those typical of natural particles, we have performed a further suite of LGG shots of spherical grains onto Stardust foil at the University of Canterbury. The projectile diameter measurements, LGG shots and crater top lip diameter measurements were performed using the same protocols as Kearsley et al., 2006. Results are plotted in Fig. 1, from which it is now possible to interpolate the relationship between crater top lip diameter and original projectile diameter for important natural species such as sulfides, mafic silicates and aluminosilicates. The larger steel sphere projectiles (88 micron average diameter, density c. 7.9 $gcm^{-3}$) used in this calibration generated craters whose top lip diameters were more than three times the foil thickness (101 microns), and were of greater depth than the foil thickness. This was confirmed by impacts on the surrounding witness plate (also Al1100) which showed depths in excess of 280 microns. As these larger craters clearly penetrated the entire foil in our light gas gun targets, resulting in detachment and uplift of the foil, they show incomplete coupling of flow and may give an anomalously narrow crater diameter in the foil. The broader top lip diameter seen in thick witness plate craters from 45 micron steel projectiles when compared to craters on 101 micron foil shows that further work is required to establish an accurate calibration for smaller high density particles. The calibration value for steel given in this paper (and plotted in Figs 2 and 3) is based solely upon the shot of projectiles with 45 micron mean diameter onto 2mm thickness Al100 witness plate, and should be taken as a minimum value.

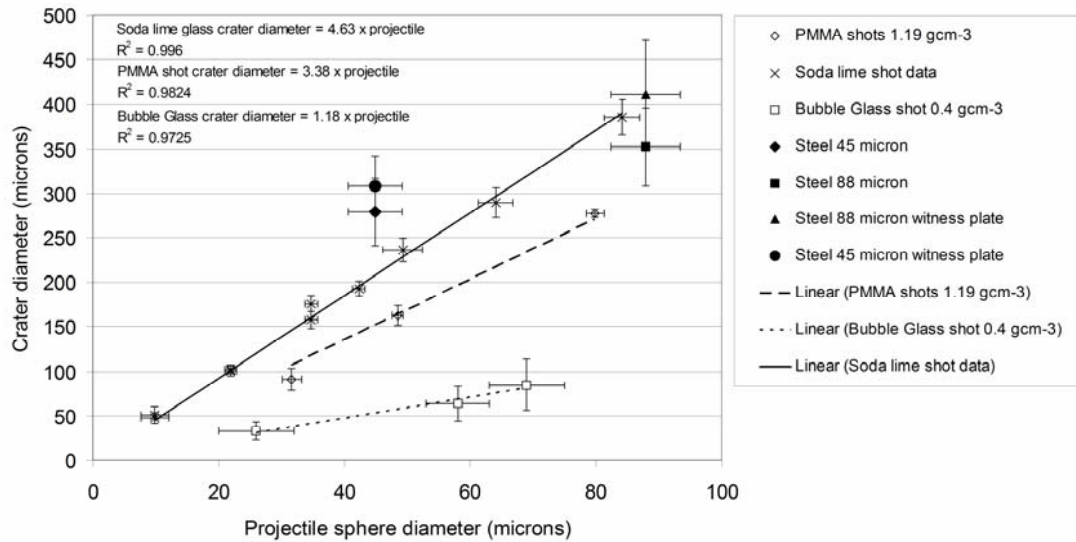

Figure 1. Calibration plot showing relationship of crater top-lip diameter (Kearsley et al., 2006) to projectile diameter, as determined for LGG shots of four materials of different density onto Al1100 Stardust foil, at velocities close to 6 kms$^{-1}$. Projectiles and craters produced by impacts of Poly methylmethacrylate, soda lime glass and steel spheres were measured by SEM; 3M Scotchlite® bubble glass spheres and their impact craters were measured by optical microscopy. Error bars are 1 standard deviation from average. Data for soda lime craters from Kearsley et al. (2006).

The original calibration of Kearsley et al. (2006), based upon impacts by soda-lime glass spheres of density c. 2.4 g cm$^{-3}$, revealed a well-defined linear trend for crater top lip diameter as a function of particle size, across a projectile size range from 9 to 84 microns. Our new experiments (Fig. 1) have demonstrated that projectile density also exerts substantial control upon crater diameter, and it is clearly important to determine density for an impactor in order to apply an appropriate calibration equation. How can impactor density be inferred for particles of unknown composition and internal structure?

**Crater morphology and depth as a function of impactor density**

Crater form from impacts on Al1100 has been described at length by Bernhard and Hörz (1995), Love et al., (1995), and in the recent Stardust foil calibration study of Kearsley et al. (2006). The introduction of software capable of generating digital elevation models (DEM) and profiles from tilted stereo pair electron micrographs now permits precise and accurate measurement of three-dimensional crater morphology. We have used the Alicona MeX® program to produce DEM for over 100 individual impact features created by particles of known composition and density, under velocity regimes and geometrical arrangement very close to those of the Stardust encounter with Wild 2. From the DEM we have generated vertical depth profiles along lines from outside the crater lip and through the deepest part of the craters, as illustrated in Fig. 2. The representative vertical depth profiles of Fig. 2 have been arranged in sequence of increasing projectile density from top to bottom of the diagram, and are scaled so that the top lips of each are aligned. No adjustment has been made in the horizontal to vertical aspect ratio of the craters, so the shape differences are real. A comparison of crater floor depth below the ambient plane (the horizontal line) clearly shows that low density impactors (such as PMMA) yield shallow craters when compared to dense projectiles such as pyrrhotite or steel. Dimensions of internal crater diameters and maximum depth were extracted by using digital callipers within the MeX® program, and these are plotted in Fig. 3.

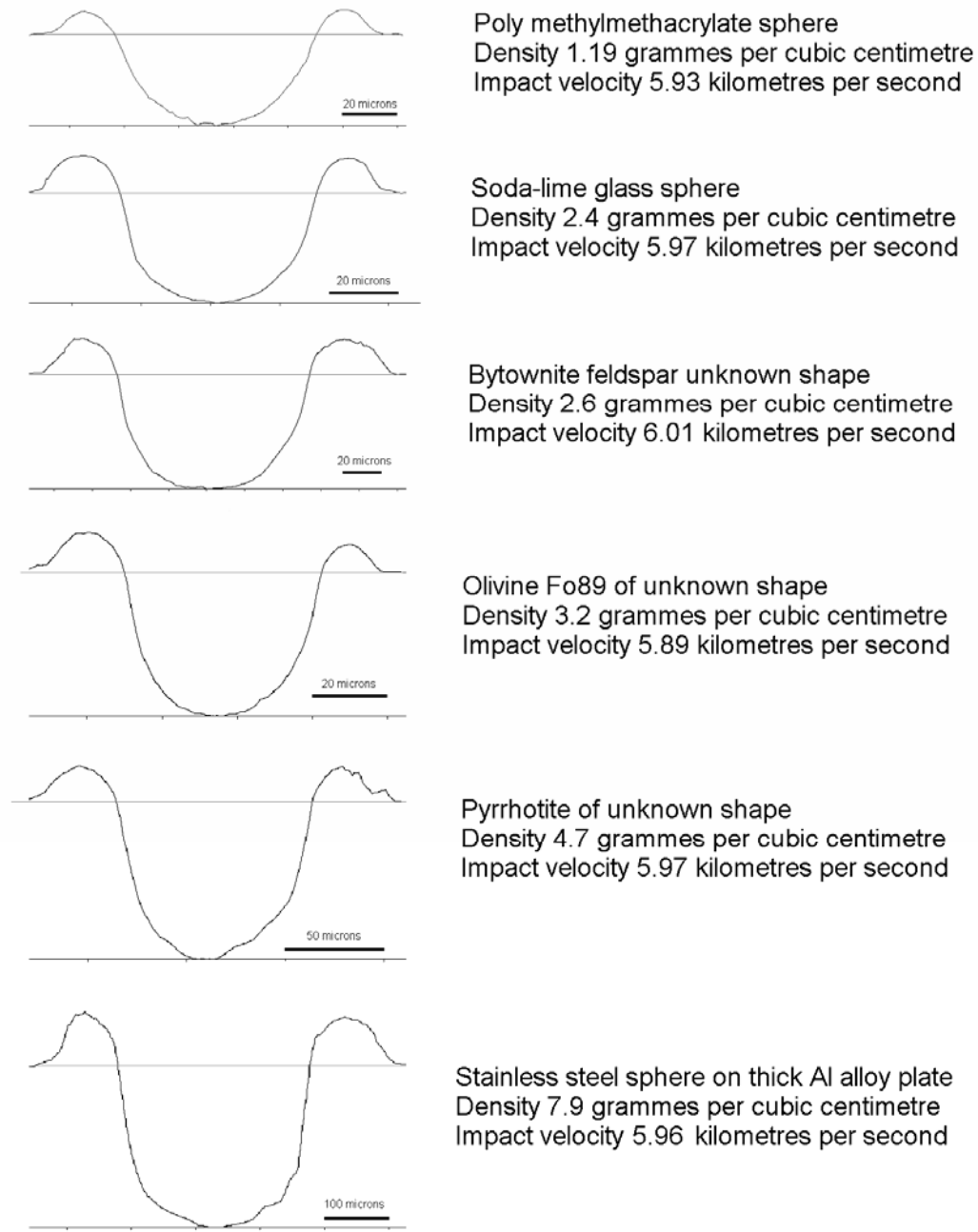

Figure 2. Typical crater depth profiles for impacts by projectiles of different densities, derived from stereo-pair electron micrograph analysis, using Alicona MeX® software. In each case, the profile in the illustration was rescaled to give the same width for the crater top-lip diameter (D of See et al., 1994; and Kearsley et al, 2006). The profile aspect ratio was kept constant, giving a true reflection of variation in crater depth. Note close approximation to 'bowl-shape' in each case. The thin horizontal line on the each crater profile shows the ambient plane (the undisturbed foil surface away from the crater), as used to measure the internal crater diameter. A vertical line drawn downward from the ambient to the lowest point on the crater floor is used as a measure of crater depth.

Burchell and Mackay (1998) noted that impacts onto aluminum of dense elongate projectiles (with a maximum : minimum dimension ratio exceeding 4) will yield craters with a wide range of crater circularity, depth and offset of deepest point. For our calibration of crater diameter as a function of projectile density we had employed near-spherical glass, polymer and steel projectiles. The SEM determinations of crater top lip diameters in Fig. 1 are the average of three diameter measurements (as in the protocol of Kearsley et al., 2006). Imagery of the mineral grains used in this study also revealed that very few exceeded max : min dimension ratio of 2, and there is little dispersion of crater circularity in craters created by the same projectile type, although the diameter: depth ratio is more variable. For the measurements summarised in the crater depth/diameter plot of Fig 3, for each crater we used a single measure of the internal ambient plane diameter, on the same line as the maximum depth point. Nevertheless, we find that the great majority of craters show little deviation from symmetrical 'bowl' shaped profiles (fitting the $4^{th}$ order Zernicke polynomial description of Wallis et al, 2002) until their depth approaches the foil thickness, when their base flattens (and their form shows an increasing proportion of $11^{th}$ order Zernicke polynomial of Wallis et al., 2002). Circular rim morphology (crater max/min diameter < 1.1) occurs even where the projectiles included cleavage-bounded rhomboidal fragments or inequant shards, and were not very close to spherical in shape. However, the detailed internal crater profile may show irregularities of slope that are probably related to projectile shape, and in other experiments we have observed that elongate mineral projectile grains (max/min dimensions >5) can indeed yield elongate boat-shaped craters. The scaling algorithm quoted by Cour Palais (1987) and Humes (1991) notes a relationship between crater depth and projectile density, and correlation between projectile density and crater depth can be seen clearly in our data (Fig. 2).

We have set out to test whether it is possible to estimate the density of an unknown impacting particle from crater morphology, for example from measures of the depth/diameter relationship. Pairs of tilted SEM images, with 6° angular difference, were taken of a wide size range of craters from buckshot firings of known mineral powders. Following stereometric analysis to derive crater morphological measurements in three dimensions from the paired SEM images, using Alicona MeX®) software at NHM, the internal crater diameter was measured at the ambient target plane (i.e. below the top-lip as used in the crater vs. particle diameter calibration), and the depth as the perpendicular from the ambient plane to the lowest point within the crater (Fig. 2). The height of ambient plane was projected from the digital elevation model of the area surrounding the crater, from a distance beyond the local annular uplift which typically extends to a distance of about 2 times the crater internal radius. To demonstrate the uniformity of shape for craters produced by a single projectile composition, a plot of depth and diameter dimensions for a wide size range of soda lime impact craters, measured on a random selection of craters from the same experimental shots as Kearsley et al. (2006), is shown in Fig. 3a (left). For craters up to 101 microns depth (i.e. the foil thickness) and 160 microns diameter, the relationship between crater internal diameter and depth appears linear, and yields a diagnostic ratio. Larger craters on 101 micron thickness Stardust foil show continued increase in depth, but at a lesser rate compared to diameter, and accompanied by a change in morphology, with a flatter crater floor. We attribute this change in crater form to the effect of the free surface at the rear side of the foil, with approach to full-thickness penetration, which is only prevented by partial mechanical coupling with the underlying aluminum support plate, in which the crater depression may continue. A similar situation is likely to have occurred on the Stardust spacecraft, with foil close above the thick, solid Al collector frame. Variation in the spacing between foil and frame may influence crater morphology and dimensions. However, the very close size match between soda lime glass impact craters on foil-wrapped aluminium plate targets and those on foils with a free rear surface, even where crater diameter exceeds 300 microns, demonstrates that there is little dispersion of crater diameter dimensions for particles with density close to natural silicate minerals, and the two size distributions are statistically indistinguishable (Kearsley et al. 2006). Larger craters from impacts by steel (density 7.9 $gcm^{-3}$) on thick Al1100 witness plates beside the target Stardust foil do show slightly broader and substantially deeper crater profiles than on the foil, demonstrating the importance of 'semi-infinite' target thickness in controlling crater morphology for these much denser impactors.

A plot of depth and diameter measurements for craters of less than 101 microns depth (averages and one standard deviation error bars in Fig. 3b), shows that increasing impacting particle density does result in greater crater depth/diameter values, although there can be considerable dispersion for craters by the same impactor. A few craters showing shape evidence for multiple, superimposed bowls were excluded from the plot. The relatively broad range of depth/diameter values for the silicate minerals may reflect variation in their particle shape, but the poor clustering of data for soda lime craters is more puzzling. It is unlikely to be due to fragmentation in the gun as there is very little dispersion of crater diameters (Fig. 1), and shots of the same projectiles onto very thin foils (which give a good proxy for impactor dimensions) show no evidence of extensive break-up for solid soda-lime spheres. Bubble glass spheres vary in their bubble content, may undergo more fragmentation in the gun, and it is therefore difficult to assign precise density to values to each impacting grain. Although we cannot yet constrain the precise depth to diameter ratio for very low density impactors (< 0.5 $gcm^{-3}$), nevertheless the data in Fig. 3b clearly demonstrate that it should be easy to recognise impacts by materials of lower density than that of polymers such as PMMA (1.19 gcm$^{-3}$).

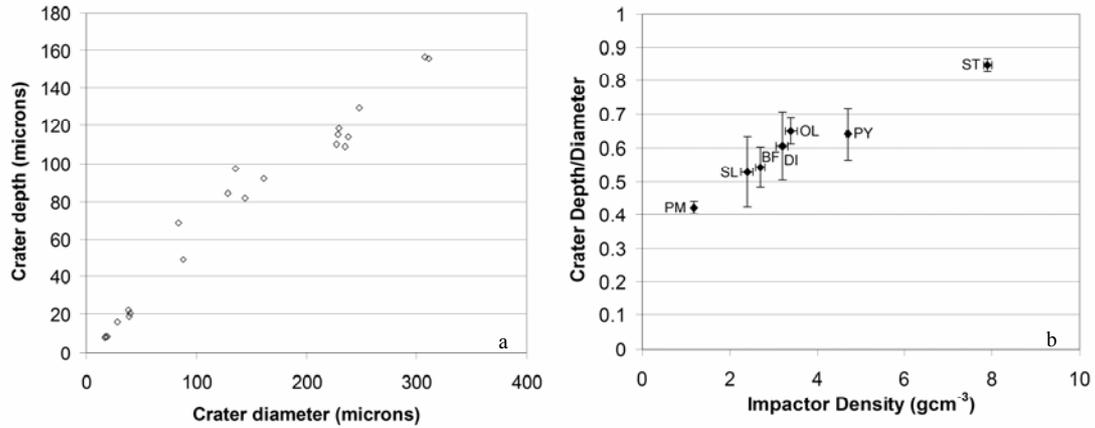

Figure 3. (a) Soda lime glass sphere impacts on Stardust foil at ca. 6 kms$^{-1}$. Crater depth to internal diameter ratio remains constant with a linear trend until the foil thickness is reached, beyond this point crater depth increases at a lesser rate. (b) Relationship between measured depth/internal diameter and projectile density for craters of less than 101 microns depth, formed by impact of : PM poly methylmethacrylate; SL soda lime glass; BF bytownite feldspar BM.2005,M312; DI diopside pyroxene BM.2005,M310; OL olivine BM.1950,337; PY pyrrhotite BM.2005,M317; ST stainless steel. All measurements are relative to the ambient target plane. Y-axis error bars are plus and minus 1 standard deviation from average depth/diameter. X-axis error bars reflect the published range of density for this material, with linear interpolation for the specific projectile composition.

From the plot of Fig. 3b, an impacting particle density can be estimated, and hence an appropriate crater diameter/ projectile diameter conversion factor can be selected from Fig.4. Together, the particle size and density determined from dimensions of a crater can yield an impactor mass.

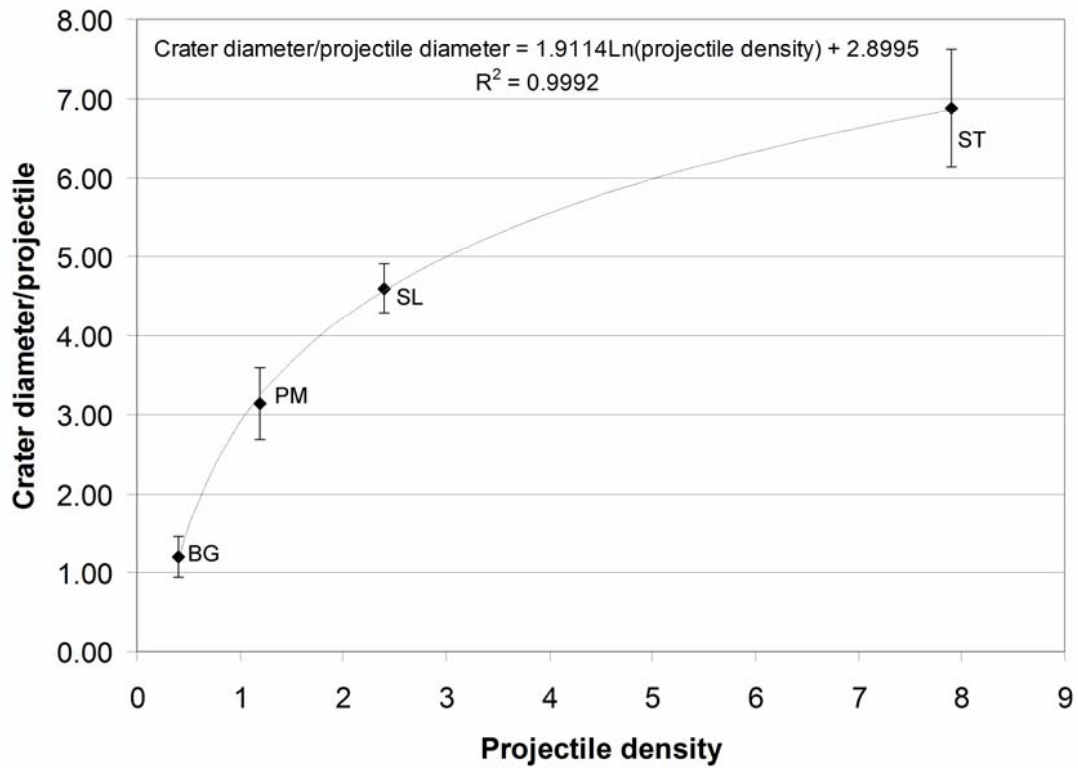

Figure 4. Relationship between impacting particle density (X axis) and the scaling factor between projectile diameter and crater top-lip diameter (Y axis). Data from calibration shots of bubble glass (BG), poly methylmethacrylate (PM), soda-lime glass (SL) and stainless steel spheres (ST). For all except stainless steel, which is based upon a single shot (see text), the crater diameter / projectile diameter (Y axis) figures are averages and one standard deviation error bars from the data plotted as linear trends in Fig. 1.

The relationship between particle density and crater top lip diameter for a particular impactor size is not linear (Fig. 1), although there is substantial increase in crater volume with density, clearly reflected in increasing crater depth compared to crater diameter (Figs 2b and 3). We have not yet obtained data for the depth/diameter ratio for the low density bubble-glass craters as we have concentrated work on those materials likely to be close analogues of the more common minerals expected to be found in cometary dust, and our observations suggest that the density of individual bubble glass particles may vary considerably, so the density of the impactor responsible for individual craters may not be well constrained. Although we were able to produce crater profiles and depth/internal diameter ratios from impacts by our pyrrhotite and olivine projectiles, as well as the monodispersive sphere samples (Fig. 3b), the mineral powders were polydispersive (with poorly constrained individual particle size). We therefore have no experimental points for these minerals on Fig. 1 and cannot give accurate figures for their crater top lip diameter/projectile diameter relationship from our current data. Nevertheless, we suggest that appropriate crater top lip/projectile diameter calibration values for minerals grains of importance to Stardust can now be inferred from Figs. 1 and 4. For example, from Fig. 4 we can deduce that an impact by an olivine grain of density 3.2 $gcm^{-3}$ might be expected to produce a crater with a top lip diameter of just over 5 times the particle diameter. If a substantial proportion of a bowl-shaped crater is preserved on Stardust foil it should therefore be possible to derive a particle density for the impactor, and from estimated particle diameter, thence a particle mass. Comparison to density inferred from compositional information will then yield an estimate of particle porosity.
Burchell and Mackay (1998) used a study of depth/diameter distribution to predict a mean impactor density of 3.4 $gcm^{-3}$ for craters found on the Long Duration Exposure Facility (LDEF) after return from low Earth orbit (LEO). However, the wide range of impact velocities and incidence angles for dust impacting on LDEF made precise interpretation difficult. By contrast, the tightly constrained velocity for dust impacting on Stardust, and the perpendicular incidence, should permit accurate determination of particle density. This is particularly important for interpretation of impactor residue composition, as internal structure may strongly influence strain concentration and post-shock peak temperature. Porous (lower density) particles are likely to experience enhanced temperatures, and consequently a greater loss of volatile components when compared to less porous particles of the same material (Shen et al., 2003). To make a reliable interpretation of a Stardust residue by direct comparison with a laboratory impact, it is therefore necessary to ensure that both impacting particles had similar internal porosity, which may be inferred by measuring their density from the depth/internal diameter ratio of their craters. Unfortunately, laboratory simulations by LGG require projectiles with high internal strength (and usually with low porosity, hence relatively high density) to survive the very rapid acceleration in the gun. This has constrained the range of suitable projectiles, and it has not yet proved possible to simulate impact by a loose, porous silicate dust aggregate of very low overall density. It is probably not appropriate to use very low density projectiles that are internally homogeneous, when the porous grains that may be present in cometary dust are likely to be loosely bound aggregates of higher density mineral and glass grains such as those seen in interplanetary dust particles. However, numerical simulations and field studies of much larger craters (e.g. Schultz and Gault, 1985; Bland and Artemieva, 2006) have shown that complex, overlapped crater fields may be created by near-synchronous impacts of a partially disrupted bolide. We believe that small impacts by loosely bound cometary dust aggregates on aluminium may also produce irregular features with a broad and shallow depth profile, but analysis of each subcomponent depression should be treated as that of an individual small crater, with consequences for residue thickness and volatile loss as explained below.

**Stardust Foil composition and limitations on residue analysis.**

The metal foil on the Stardust collector frame was designed to act as a sleeve to the side of the aerogel blocks, enabling safe emplacement within the frame and subsequent removal on the completion of the mission. It was not intended primarily as a dedicated collection substrate, and the important functional requirements were met by Al1100 alloy foil. A generic description of Al1100 (Davis, 1998) gives a content of Al greater than 99% by weight, with the majority of the rest composed of iron (Fe) and silicon (Si). Kearsley et al. (2006) demonstrated that the Fe content of the alloy is non-uniform at the micron scale, with highly localised Fe-rich areas being visible in backscattered electron images (BEI) of the foil, showing an irregular spatial distribution at the 100 micron scale, but likely to be present within any area subject to impact. There is thus a potential problem for determination of Fe content in impact residues on Stardust foil as the incorporation of Fe from alloy inclusions may cause overestimation of abundance. Our WDS and EDS X-ray maps also reveal not only widespread concentrations of Si, but also titanium (Ti) and vanadium (V), albeit restricted to small patches (Figs. 5 and 6). Additionally, ICP analysis of unexposed foil samples from the spacecraft reveals a wide range of trace element contents (Table 1) which need to be considered as potential substrate interference for sensitive bulk composition determinations of residue. Foil analyses were performed at NHM. For compositional analysis, four spacecraft foil samples, each weighing approximately 300mg, were supplied by NASA, having been removed from the collector frame after flight. The samples had not been exposed to dust impact during the mission, but were contiguous to impacted areas. Approximately one third of each foil was removed by tearing, without use of metal tools so as to avoid contamination. The samples of approximately 100 mg were cleaned in an ultrasonic bath with acetone and then with deionised water, dried and weighed. Samples were placed in acid-cleaned polypropylene sample tubes and dissolved in a mixture of 1ml HCl, 2ml $HNO_3$, 0.5ml $H_2SO_4$ in 25 ml water (an acid mixture of the type used in industrial aluminum analysis, hydrochloric acid alone would be effective for dissolution, but may not retain some trace elements in solution). The sealed tubes were warmed to 60ºC in an ultrasonic bath to speed dissolution. When dissolution was complete, the solutions were made up to 50ml with water. A process blank was prepared along with the samples. Analysis by ICPAES used a Varian Vista Pro axial instrument with simultaneous CCD detector and argon purged spectrometer. Data acquisition for each solution consisted of five replicates each of ten seconds integration period. Calibration was via multi-element standard solutions prepared from certified single element standard solutions with the addition of 2000ppm Al to match the sample matrix.  Emission wavelengths were chosen on the basis of sensitivity and lack of possible interferences. Data are reported as ppm element in the solid sample.

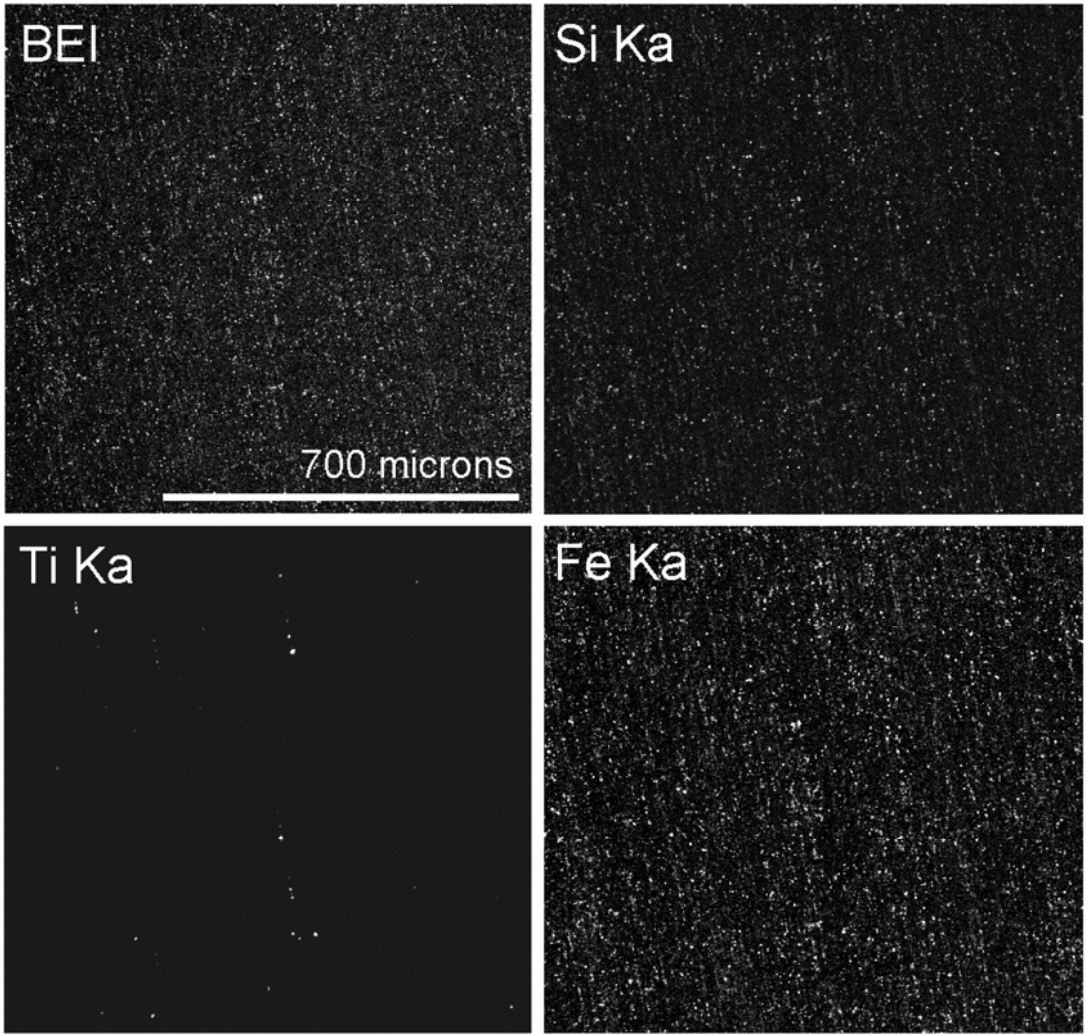

Figure 5. Foil used as a target for laboratory light gas gun impacts, backscattered electron image (BEI), and WDS X-ray maps of same area.

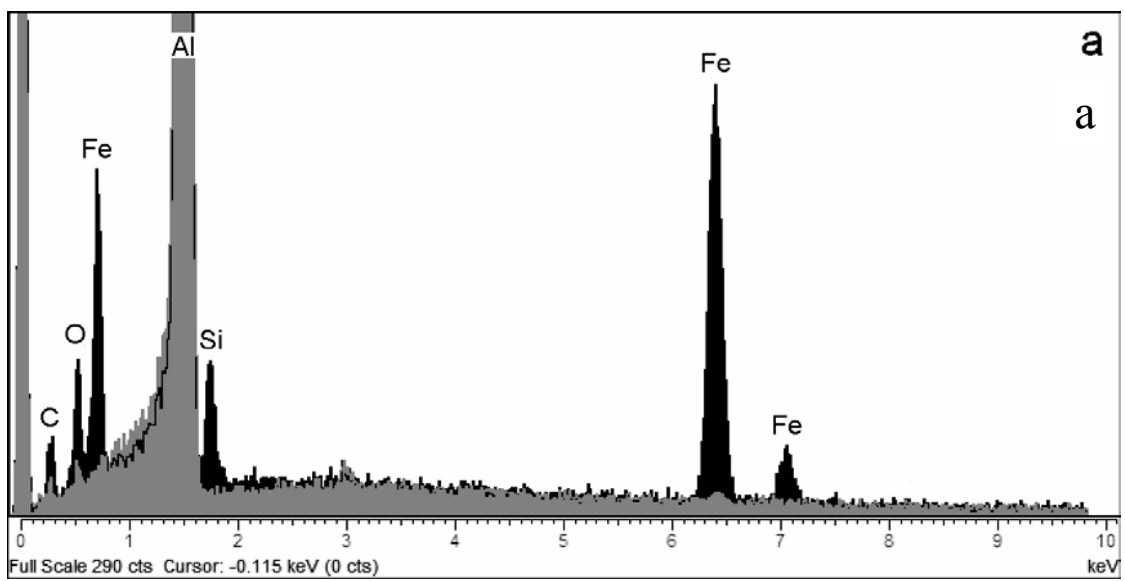

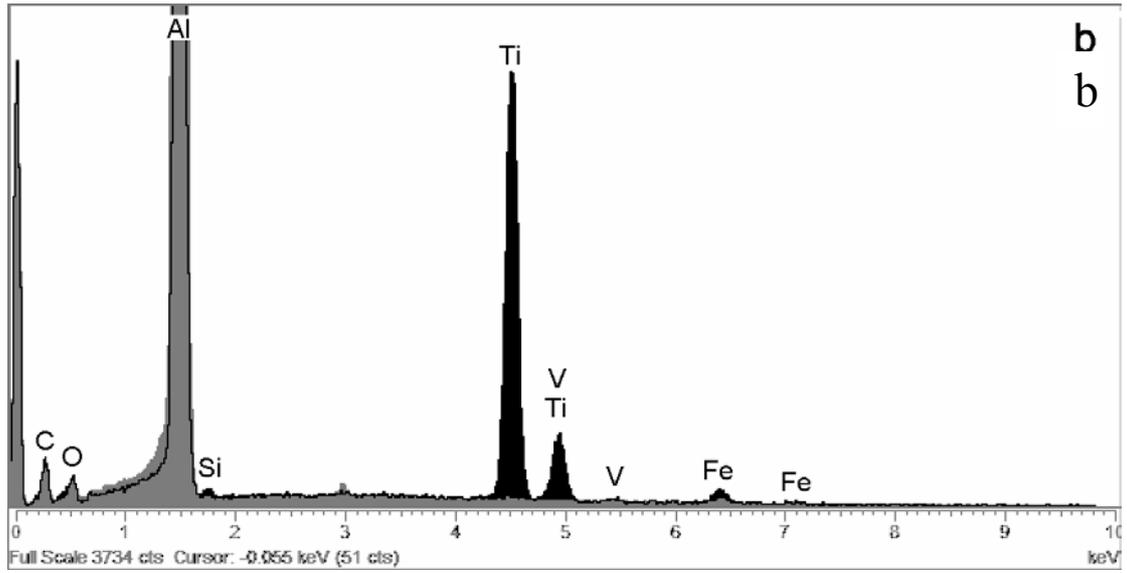

Figure 6. EDS X-ray spectra of (a) Fe-Si and (b) Ti-V rich alloy inclusions in Al1100 target foil, similar to the foils flown on the Stardust spacecraft.

Table 1. Trace element composition of Al1100 foil used on the Stardust spacecraft.

| Sample<br>Wt (g)<br>Element | CO13N,0<br>0.11316<br>ppm | CO48N,0<br>0.11143<br>ppm | CO91N,0<br>0.10128<br>ppm | C118N,0<br>0.10302<br>ppm | Average<br>ppm | St. Dev.<br>ppm | Detection<br>limit<br>ppm |
|---|---|---|---|---|---|---|---|
| Na | < | < | < | < | | | 2.5 |
| Mg | < | < | < | < | | | 0.5 |
| Si | 284 | 308 | 292 | 294 | 294.4 | 10.0 | 0.5 |
| P  | < | < | < | < | | | 25 |
| K  | < | < | < | < | | | 5 |
| Ca | < | < | < | < | | | 0.5 |
| V  | 93 | 93 | 92 | 93 | 92.6 | 0.4 | 2.5 |
| Ti | 190 | 190 | 185 | 186 | 187.8 | 3.0 | 0.5 |
| Cr | < | < | < | < | | | 0.5 |
| Mn | 20 | 20 | 20 | 20 | 20.0 | | 0.5 |
| Fe | 3575 | 3570 | 3523 | 3556 | 3556 | 23.3 | 0.5 |
| Co | 1.4 | 1.4 | 1.3 | 1.3 | 1.3 | 0.1 | 1 |
| Ni | 45 | 45 | 45 | 45 | 44.8 | | 2.5 |
| Cu | 31 | 31 | 30 | 30 | 30.8 | 0.5 | 0.5 |
| Zn | 74 | 74 | 73 | 73 | 73.4 | 0.8 | 0.5 |
| As | < | < | < | < | | | 5 |
| Cd | < | < | < | < | | | 0.5 |
| Sr | < | < | < | < | | | 0.5 |
| Ba | < | < | < | < | | | 2.5 |
| Pb | < | < | < | < | | | 25 |

Note: Al (> 99 wt%) is not included in this table. The < symbol signifies that this element was below detection limit, as given in the final column. Sample nomenclature follows the convention for location on the collector grid, as used throughout the Preliminary Evaluation of Stardust materials.

**Analysis of impact residue by energy dispersive X-ray spectrometry**

Although a wide range of microanalytical techniques are now available (Zolensky et al., 2000), EDS is usually the first elemental analysis method applied to hypervelocity impact crater residues as it is essentially non-destructive (Graham et al., 2001), although deposition of a contaminant layer upon the substrate may hinder subsequent infra-red, laser Raman spectroscopic and ion beam methods. Nevertheless, the technique has a long history of successful application to impact craters on spacecraft surfaces exposed in low Earth orbit (e.g. Bernhard et al. 1994b; Brownlee et al., 1994; Graham et al., 1999), where it has been extensively employed for the distinction of micrometeoroid and space debris impactors (e.g. Graham et al., 1999; Kearsley et al., 2005). The relative insensitivity to beam-sample-detector geometry makes EDS especially useful as it can yield some compositional information even from within complex-shaped depressions such as craters in metals, an advantage over the more sensitive and accurate WDS and time-of-flight mass spectrometers (ToF-SIMS). Unfortunately, EDS detection limits are relatively poor, and quantities of less than 0.2 weight % of many elements often cannot be determined reliably.

Modern electron beam instruments, especially those utilising a field emission electron source, allow excellent spatial resolution at high beam current and low accelerating voltage (e.g. 5 kV), sufficient to image sub-micron craters, and to excite primary X-rays that are restricted to a very shallow residue layer (less than 500nm), with a flux adequate for an EDS spectrum with well-defined characteristic peaks and little contribution from the substrate below. Such spectra quickly yield a diagnostic check-list of most of the major elements present in a crater, but are unlikely to show minor concentrations of elements heavier than calcium (Ca) as the beam does not have sufficient energy to excite their K lines, and the L lines are heavily overlapped.

In this paper we concentrate on evaluation of quantitative analysis in craters of over ten microns diameter. Here, relatively deep 'sampling' of X-ray emission below the sample surface (from several microns depth if an accelerating voltage of 20 kiloVolts is used) may permit recognition of thicker (micron scale) residue areas, less likely to have undergone major elemental fractionation (see below). The high energy beam also provides excitation of higher energy K lines from Ti, chromium (Cr), manganese (Mn), Fe and nickel (Ni). The substantial contribution of Al Kα X-rays in spectra from crater residues might be expected to cause difficulties in peak and background fitting (and hence quantitation), due to a major step in the Brehmstrahlung (background) radiation at the Al Kα absorption edge. In practice this is often not a problem because Al is not in a location able to absorb the higher energy X-rays, unless it is in the path from the spectrum acquisition point to the X-ray detector. However, Kearsley et al. (2006) have demonstrated in FIB sections that interlayering of residue and substrate metal may occur on submicron scales, and it is therefore not possible to obtain reliable quantitative Al analyses in situ within craters. When the Al absorption edge is seen to be substantial, i.e. an analysis is being taken from an area of shadowing, and if Al is included in the list of elements to be determined, the step and peak may also distort the X-ray matrix correction for other elements. As a consequence, in this paper we exclude Al Kα from spectrum fitting of all quantitative analyses taken in situ from thinner residues in craters where there is evidence of absorption. Unfortunately, this excludes a large proportion of craters of less than 10 microns diameter. However, good Al data can be obtained from FIB TEM ultrathin sections taken from very small impacts of basalt glass, usually with little or no interference from the metal substrate.

Where we perform in situ quantitative analysis of thicker residues and see no evidence of absorption, we do include Al in the peak search and matrix correction, although in our summary we discount it and normalise our data for the other elements. When we consider the effects of sample geometry and elemental fractionation from the impact residue, we plot characteristic X-ray peak areas rather than element quantities, and include Al Kα, especially as it gives a proxy for residue thickness. Display of the spectrum 'fit' in the INCA software shows an excellent match of peaks and Brehmstrahlung if Al Kα is included, and we are therefore confident as to the accuracy of the peak area measurements.

In studies of LEO impacts, an expectation of major elemental fractionation at the higher pressures and temperatures implicit in micrometeoroid velocities ($> 20$ kms$^{-1}$), and the uncertainty as to individual particle impact velocity has made detailed interpretation of residue chemistry impracticable. However, the nature of the Stardust encounter yielded a constant and relatively low velocity for all impacts on the spacecraft (6.1 kms$^{-1}$ Brownlee et al., 2003). At this velocity, impacts of silicate minerals onto aluminum alloy should yield peak pressures less than 100GPa (Stöffler, 1982; Bernhard et al., 1994b), and should not result in major loss of vapour. Rapid quenching of residue is aided by the very short duration of the crater forming process (microseconds), and excellent thermal conductivity of the metal substrate. There is thus much greater potential for retention of recognisable impactor chemistry on Stardust compared to earth-orbiting collectors.

We have used two approaches for analysis of residues: collection of EDS X-ray spectra from the interior of intact craters created by impacts of over twenty different mineral materials; and FIB sectioning of small craters, followed by EDS analysis in an analytical TEM. The former approach is very rapid, with a useful 'check-list' of element characteristic X-ray peaks being acquired from unprepared samples in a few tens of seconds. Unfortunately, there are many difficult complications in matrix correction and consequently only limited quantification is possible, especially for very small craters. FIB-TEM yields high quality results from even sub-micron craters, requiring no matrix corrections and thereby making quantitative data both much easier to interpret and more reliable. The preparation of ultra-thin slices of crater residues using FIB techniques can be time-consuming, as a result only a limited number of sections may be produced during the preliminary evaluation of Stardust foils. Also, the preparation of electron transparent sections by sputtering does result in the removal and sometimes partial re-deposition of material, therefore it may not be suitable for application to every crater. However, the excellent spatial resolution possible with analytical TEM of ultra-thin sections can reveal diverse co-existing residues at sub-micron scale, and allows inclusions within the alloy to be avoided. The ability to distinguish crystalline materials from amorphous residue by dark field or diffraction imagery, and the precision (reproducibility) of FIB-TEM EDS compositional information also make this the technique of choice for residue determination in small craters.

To unravel the most significant complications that make interpretation of EDS data from craters difficult we have considered: 1) geometrical issues of X-ray generation in the residue, absorption in residue and crater, and collection at the detector, all of which may affect in-situ EDS analysis; and 2) the change in chemical composition of the residue as a result of impact, observed in intact craters, and quantified in FIB-TEM analyses of ultrathin sections.

We have utilised light gas gun shots of a wide range of mineral materials to test our models and techniques. The detailed results from our study of diverse mineral species will be published elsewhere (Wozniakiewicz et al, in prep). In this paper we only consider analysis of craters produced by impacts of olivine (Admire BM.1950,337), diopside (BM.2005,M310), basalt glass (USGS NKT-1G) and pyrrhotite (BM.2005,M317).

**Geometrical effects on EDS analysis.**

An electron beam focussed onto the surface of a residue-lined impact crater will be able to excite X-rays from any location that the beam can reach. Unfortunately, the conventional inclined position for the EDS detector prevents collection of X-rays directly from the floor of a typical crater (e.g. the olivine impact shown in Fig. 4 above), as the crater wall obscures the 'line of sight' (Fig. 7). If the foil surface is kept perpendicular to the electron beam axis, this will restrict unobstructed X-ray collection to a band around the crater lip and on the steep crater wall, facing toward the detector. Although a high X-ray flux can be collected from the crater wall, the oblique beam to sample incidence gives X-ray excitation from a very shallow volume, and enhanced low energy X-ray escape from the surface. As a result, the count rate ratios for elements such as sodium (Na) and magnesium (Mg) are increased relative to silicon (Si), making quantitative matrix correction and direct comparison to quantitative calibration standards difficult. This effect can be demonstrated both by modelling X-ray generation and escape from an inclined surface using the extended Pichou and Pouchoir (XPP) software in the INCA package, and in spectra collected from real impact craters. If the foil is tilted directly toward the EDS detector, at an angle specific for the geometry of the particular analytical instrument, it is possible to collect X-rays emitted from the crater wall to floor arc, with beam–normal incidence on the residue. Under these circumstances, quantitative matrix corrections become applicable as for everyday quantitative EDS electron microprobe analysis, although one should expect poorer precision from the rough residue surface and lack of conductive carbon coating (as is seen in Tables 2, 3 and Fig. 9). It is also necessary to normalise the wt % data for elements due to exclusion of aluminum from the total. Our experiments with laboratory impacts of olivine and diopside grains of known composition (Figs. 8, 9 and table 2) show that this tilted orientation gives analytical results with stoichiometry closer to that of the projectile material, implying little or no compositional change, confirmed by analysis of ultrathin sections.

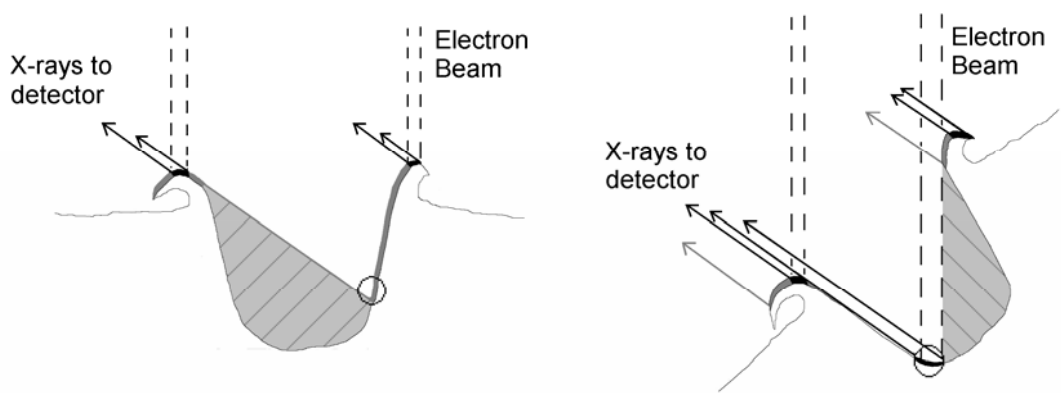

Figure 7. X-ray generation in an impact crater of typical depth profile, based upon an impact by olivine, and their collection by an EDS detector, shown for foil that is beam-normal (left) and tilted (right). The electron beam is shown as the dashed lines, areas that yield detectable X-rays in grey (difficult matrix corrections) or black (reliable matrix corrections, giving good stoichiometric analyses). Residue on the lip is thin and sparse, although residue is thicker and common close to the bottom of the crater walls. Cross-hatched areas yield little or no useful X-ray data for analysis, circled areas are locations where the thickest residue is usually found.

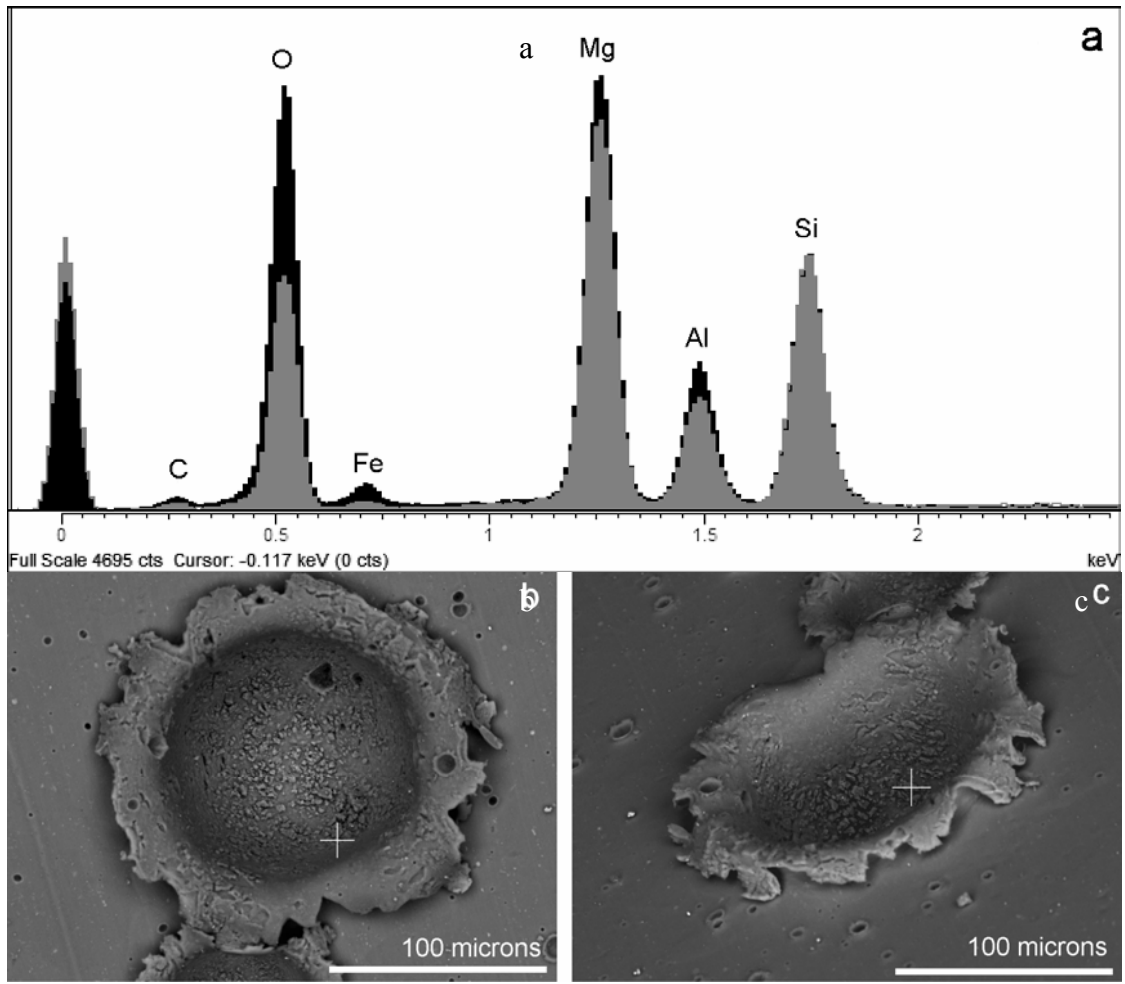

Figure 8. Olivine LGG impact. a) two EDS spectra from the same crater, normalised to Si counts. The black spectrum was taken with foil perpendicular to the beam (c. 65° beam incidence angle onto inclined residue on the crater wall in b), grey spectrum from same residue, with foil tilted toward the detector at 40 degrees, giving beam-normal incidence on crater floor c).

Table 2. Olivine (BM.1950,337) LGG impacts, comparison of analytical results from a polished section of olivine WDS analyses, tilted (beam-residue normal) and foil-normal (beam-residue oblique) crater surfaces by EDS.

|  | Mg | Si | Mn | Fe | O |
|---|---|---|---|---|---|
| Wt % polished section WDS (Avg.) | 29.30 | 18.99 | 0.20 | 8.83 | 43.56 |
| 100 analyses (St. dev.) | 0.12 | 0.12 | 0.01 | 0.10 | 0.18 |
| Wt % crater residue EDS tilt (Avg.) | 28.9 | 19.4 | 0.2 | 8.7 | 43.7 |
| 5 analyses (St.dev.) | 0.7 | 0.3 | 0.1 | 1.0 | 0.7 |
| Stoichiometry in WDS section | 10.58 | 5.96 | 0.03 | 1.44 | 24.00 |
| Stoichiometry (tilted, EDS) | 10.72 | 5.95 | 0.03 | 1.34 | 24.00 |
| Stoichiometry (foil normal, EDS) | 10.96 | 5.96 | 0.02 | 1.09 | 24.00 |

From Table 2 it is apparent that the tilted foil (beam-residue normal) analyses give substantially closer match to the stoichiometry of known projectile composition. For larger craters, with a top lip diameter in excess of 20 microns, and with suitable tilted geometry, it is usually possible to find a location for analysis. Although EDS data from beam-normal foil orientation may be very useful for location of residue, e.g. by mapping, as in Fig. 8, our results and models show there is a consistent over-estimation of light elements and these spectra should not be used for accurate stoichiometric determination of residue. However, other mineral shots of refractory minerals do show that foil-tilted spectra may give quantitative analyses of major elements that are close to the known projectile composition, although with relatively poor precision (Table 3 and Fig. 9).

Table 3. Diopside LGG impacts, comparison of EDS results from 12 analyses of polished section of diopside (BM.2005,M310), and 5 spectra from tilted (beam-residue normal) crater surfaces, impact on Stardust foil at 6.01 kms$^{-1}$.

|         | Weight % | Na  | Mg  | Si   | Ca   | Mn   | Fe  | O    |
|---------|----------|-----|-----|------|------|------|-----|------|
| Section | Average  | 0.4 | 7.6 | 24.7 | 16.5 | 0.3  | 7.9 | 42.5 |
|         | St. dev. | 0.1 | 0.1 | 0.2  | 0.2  | <0.1 | 0.2 | 0.3  |
| Crater  | S3sp1    | 0.3 | 7.8 | 24.8 | 16.7 | 0.4  | 7.4 | 42.4 |
|         | S3sp2    | 0.4 | 7.9 | 26.5 | 14.8 | 0.2  | 6.7 | 43.5 |
|         | S3sp     | 0.3 | 6.7 | 23.3 | 14.0 | 0.3  | 6.2 | 49.2 |
|         | S3sp4    | 0.4 | 9.1 | 23.2 | 16.5 | 0.4  | 8.6 | 41.7 |
|         | S4sp2    | 0.5 | 7.7 | 26.0 | 15.1 | 0.2  | 7.2 | 43.1 |
|         | average  | 0.4 | 7.9 | 24.8 | 15.4 | 0.3  | 7.2 | 44.0 |
|         | St. dev. | 0.1 | 0.9 | 1.5  | 1.2  | 0.1  | 0.9 | 3.0  |

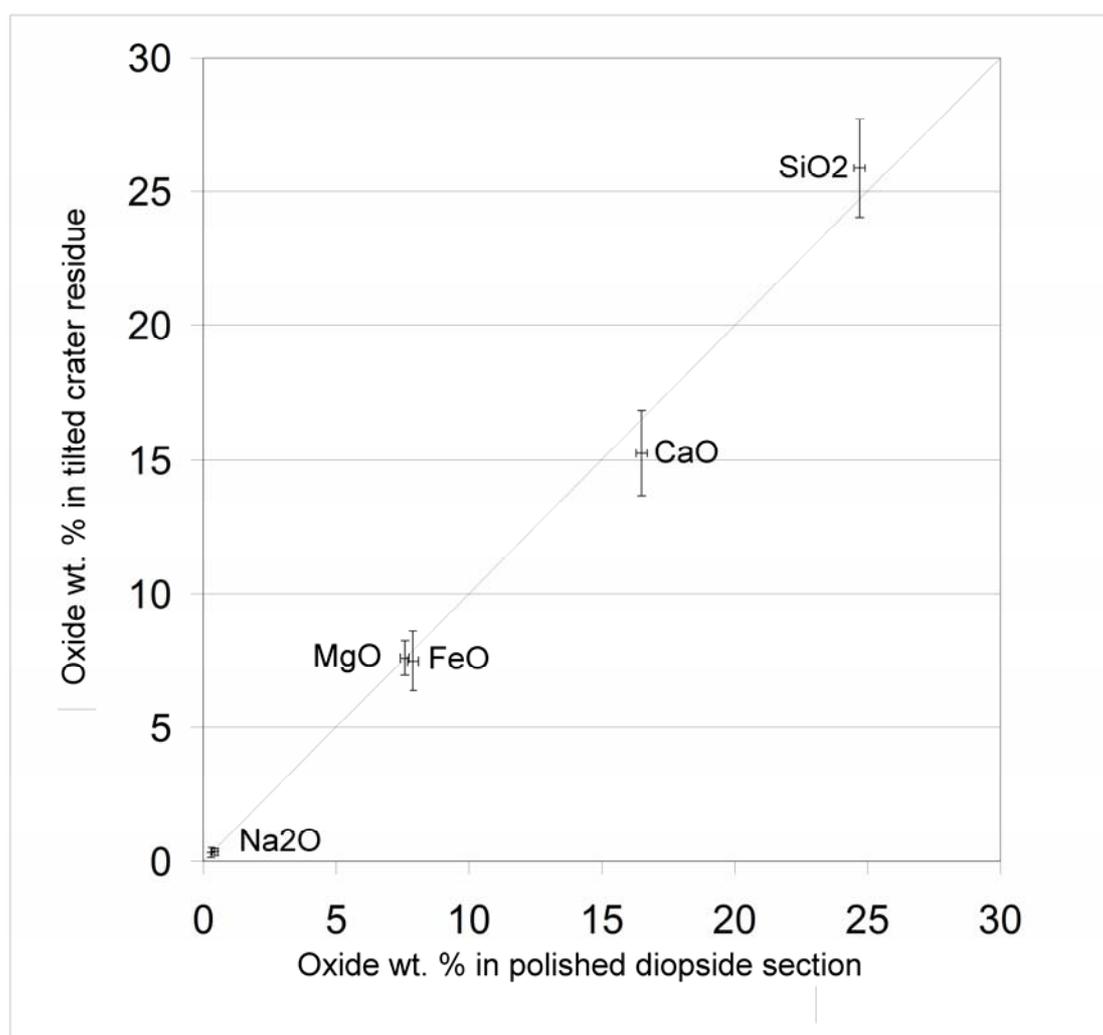

Figure 9. Plot of element weight % oxides, comparing projectile composition from a polished section of diopside BM.2005,M310 (average and standard deviation from 12 analyses) with a rough residue on the floor of a foil-tilted crater created by impact of diopside on Al1100 foil at 6.01 kms$^{-1}$ (average and standard deviation from 19 analyses). All analyses by EDS.

A second geometrical effect may cause even more problems: obstruction to the X-ray detector line of sight by the crater wall, particularly a problem for foil-beam normal incidence. This is particularly well displayed in the sulfur (S) and Fe peaks in spectra taken from impacts by the Fe sulfide pyrrhotite (sample BM.2005,M317).

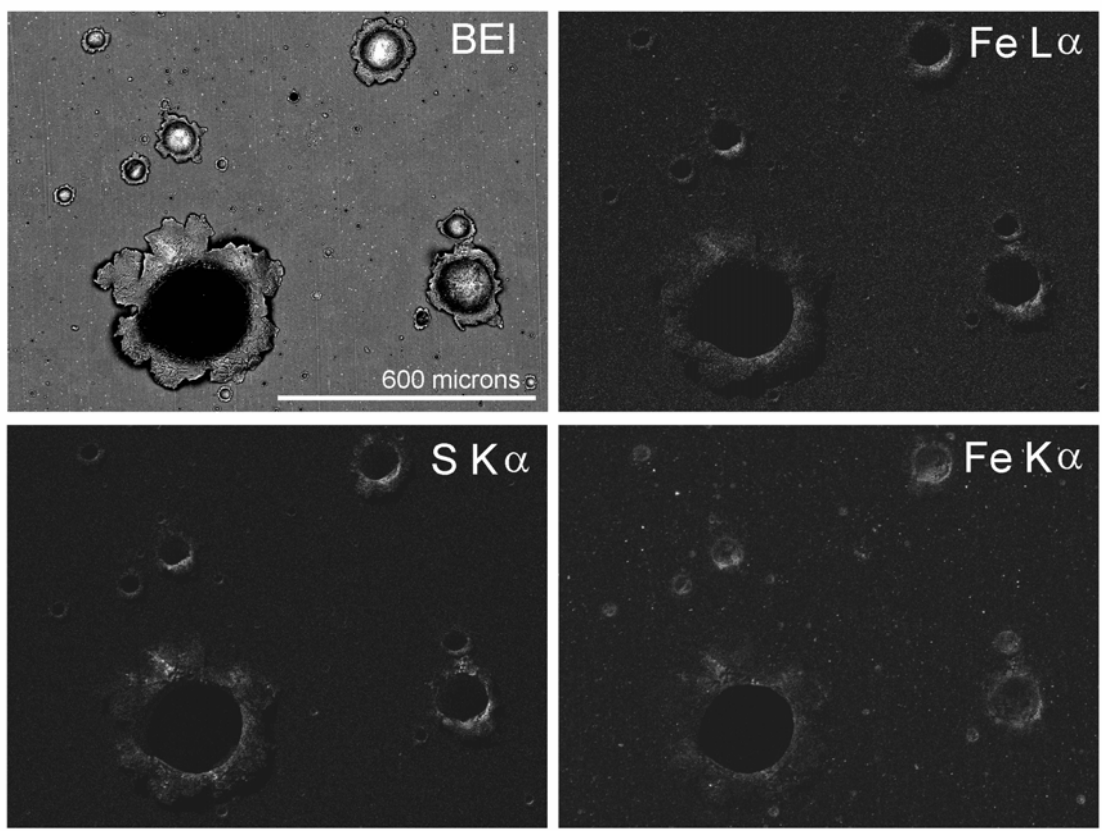

Figure 10. Light gas gun impacts by pyrrhotite (BM.2005,317) projectiles. Backscattered electron image, X-ray maps for S Kα, Fe Kα and FeLα. Beam accelerating voltage was 20kV. Electron beam normal to foil surface (i.e. no tilt).

Fig. 10 shows a strong 'shadow' effect for the lower energy X-ray lines (S Kα and Fe Lα) from the crater floor, due to absorption by the dense Al of the crater wall, obscuring the line of sight to the X-ray detector. Collection of some Fe Kα X-rays is possible even from within the 'shadow' due to the greater penetrating power of their relatively high energy. If the Fe Lα and S Kα peaks are used for analysis (as is necessary at low beam accelerating voltages e.g. 5kV), the extreme attenuation of Fe Lα in the 'shadow' area would give a false impression of low Fe:S ratio. In areas of the crater outside the shadow but on a sloping wall, the shallow interaction volume will boost Fe Lα escape, and thus overestimate Fe:S. Given the depth of interaction volume (c. 0.5 microns for Al Kα generation in aluminum at 5kV), in a micron scale crater it is likely that every electron beam spot location will encounter a different combination of geometrical effects. When a high accelerating voltage (e.g. 20kV) is employed, Fe Kα may be used for analysis, but the greater absorption of S Kα in the 'shadow' area now creates a false impression of high Fe:S ratio. With a high accelerating voltage it is possible to generate sufficient Fe Kα flux for comparison of K and L count rates across the crater, and locations with substantial X-ray absorption due to shadowing can be recognised by high K/L counts (e.g. Fig. 15 below). Another useful indicator of 'shadowing' is found in the Brehmstrahlung gradient across the segment of the background spectrum above the Al Kα X-ray line of all spectra without high phosphorus (P) or chlorine (Cl) content, between 1.8 keV and 2.8 keV (Table 4). A negative gradient, i.e. decline in background as energy increases, suggests little absorption by Al, and is the usual condition from an unobstructed view of a surface. A positive gradient (i.e. the background rises as energy increases above Al Kα) indicates substantial X-ray absorption close to the Al K edge. Although lower accelerating voltages may yield spectra from a shallower residue thickness and hence contain a greater proportion of X-rays from the residue instead of the underlying Al, they also suffer a similar problem of absorption in craters, and may not have sufficient X-ray counts to reveal higher energy lines. Recognition of absorption by shadowing must then rely on the Brehmstrahlung gradient alone, made difficult if count rates are low.

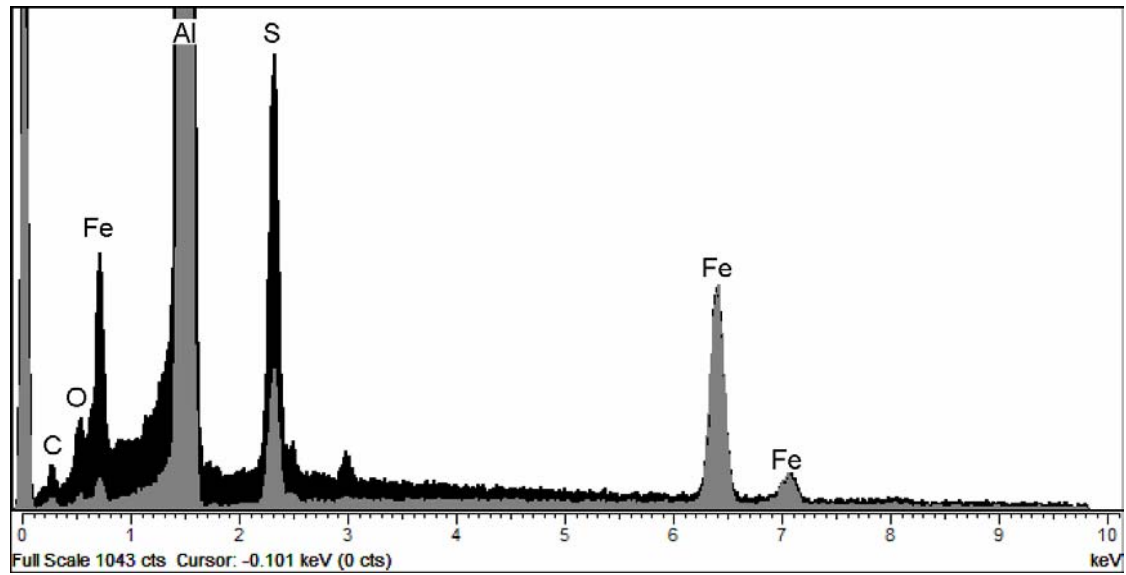

Figure 11. Pyrrhotite LGG crater, EDS with linear scale. Projectile BM.2005,M317, impacted at 5.97 kms$^{-1}$ onto Al1100 foil. Absorption of low energy X-rays by crater wall obscures line-of-sight to the X-ray detector. The grey spectrum is 'in shadow', black spectrum 'out of shadow'. Spectra normalised to Fe K$\alpha$ maximum.

Table 4. Impact residues from LLG shot of Pyrrhotite projectiles (BM.2005, 317)

|  | crater wall facing detector | crater floor in 'shadow' | Pyrrhotite section beam normal | XPP model 70° slope | XPP model 40° slope |
|---|---|---|---|---|---|
| FeL/FeK | 0.62 | 0.08 | 0.17 | 0.41 | 0.56 |
| S/Fe K | 1.46 | 0.43 | 1.78 | 3.25 | 3.79 |
| S/Fe L | 2.36 | 5.67 | 10.52 | 7.83 | 6.79 |
| Brehm. gradient | -3.10 | 11.50 | -12.40 | -4.40 | -8.70 |

X-ray count ratios for different crater locations, polished section and extended Pichou and Pouchoir (XPP) simulations of inclined surfaces relative to the electron beam.

A very low figure for Fe L/Fe K indicates strong absorption, as does a positive value for the Brehmstrahlung gradient determined by the subtraction of the 2.6 – 2.8 keV count rate from that between 1.8 – 2.0 keV. In larger craters it is relatively easy to find a location from which there is a good line of sight to the X-ray detector. It will still be necessary to tilt the surface toward the detector to give beam to residue normal incidence, or S/Fe K will be anomalously high due to low absorption, as is seen in the XPP model data above. The lack of a conductive carbon coating, and the inevitable roughness of the residue surface also introduce factors that alter the matrix correction (especially X-ray absorption) and make comparison to standards difficult. In smaller craters it is much more difficult to find either a location 'out of shadow' or with the correct beam-specimen incidence angle.

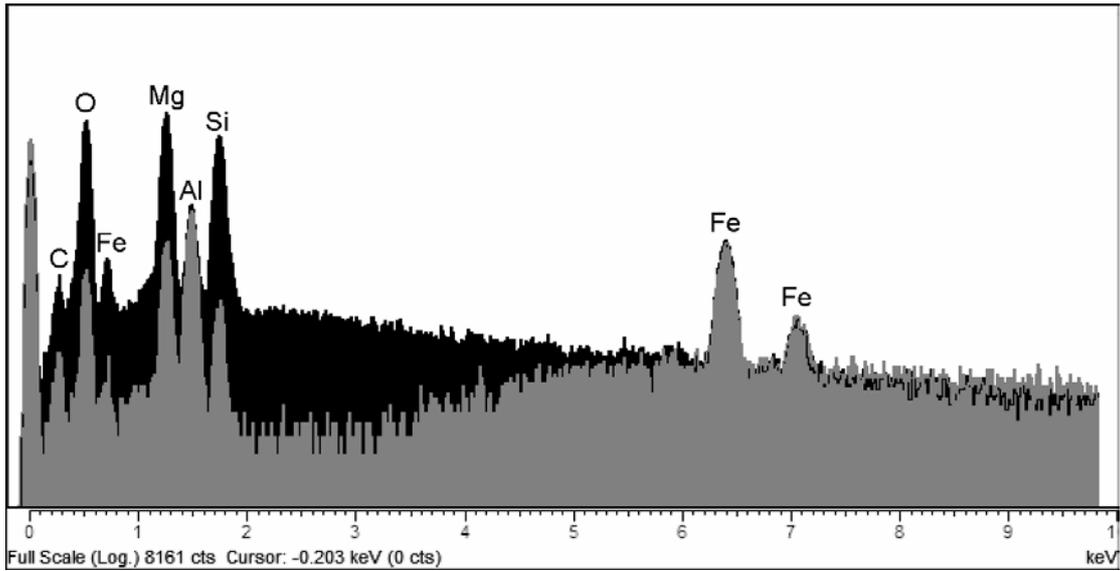

Figure 12. Residues in olivine (BM.1950,337) LGG crater, impacted at 6.07 kms$^{-1}$ onto Al1100. EDS with logarithmic scale. Absorption of low energy X-rays by the crater wall shown by the grey spectrum 'in shadow', black spectrum 'out of shadow'. Spectra normalised to FeKα maximum.

Unfortunately, the 'shadow' effect poses a particularly severe analysis problem for small craters (less than 20 microns diameter), especially as the very thin residue sheet yields a low X-ray flux and permits beam penetration into the substrate. As a consequence, it is not possible to obtain consistent elemental ratios from small craters, as is demonstrated by the spread of values for basalt glass impact plots in Fig. 13 and the pyrrhotite impacts in Fig. 15.

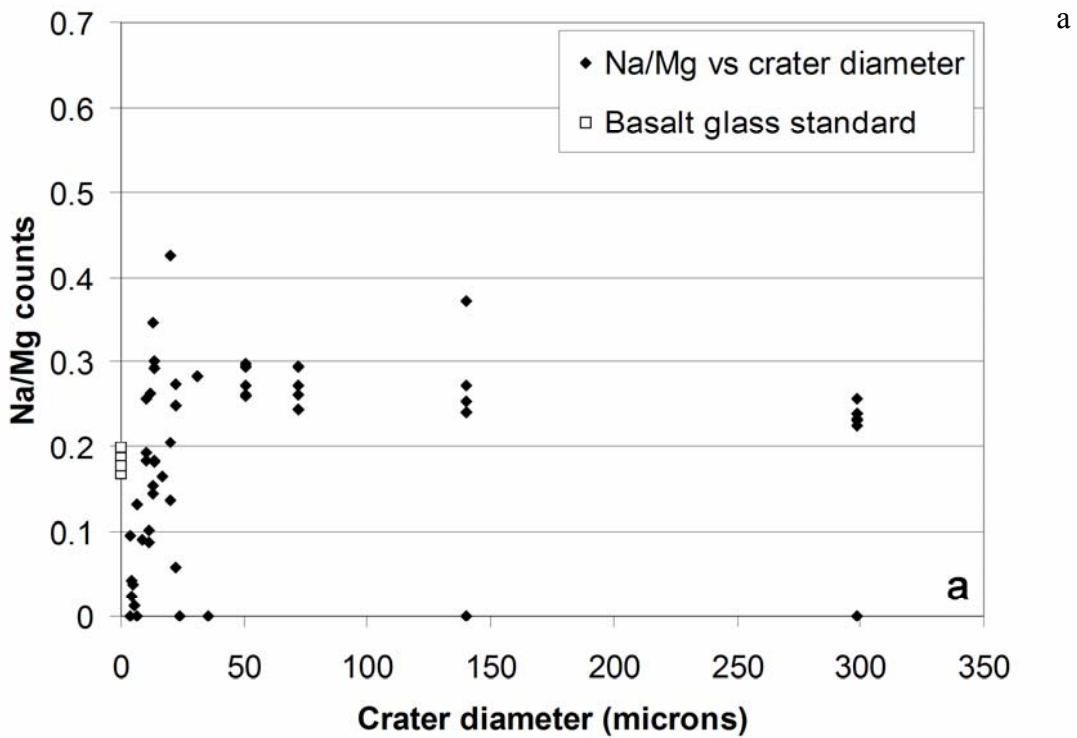

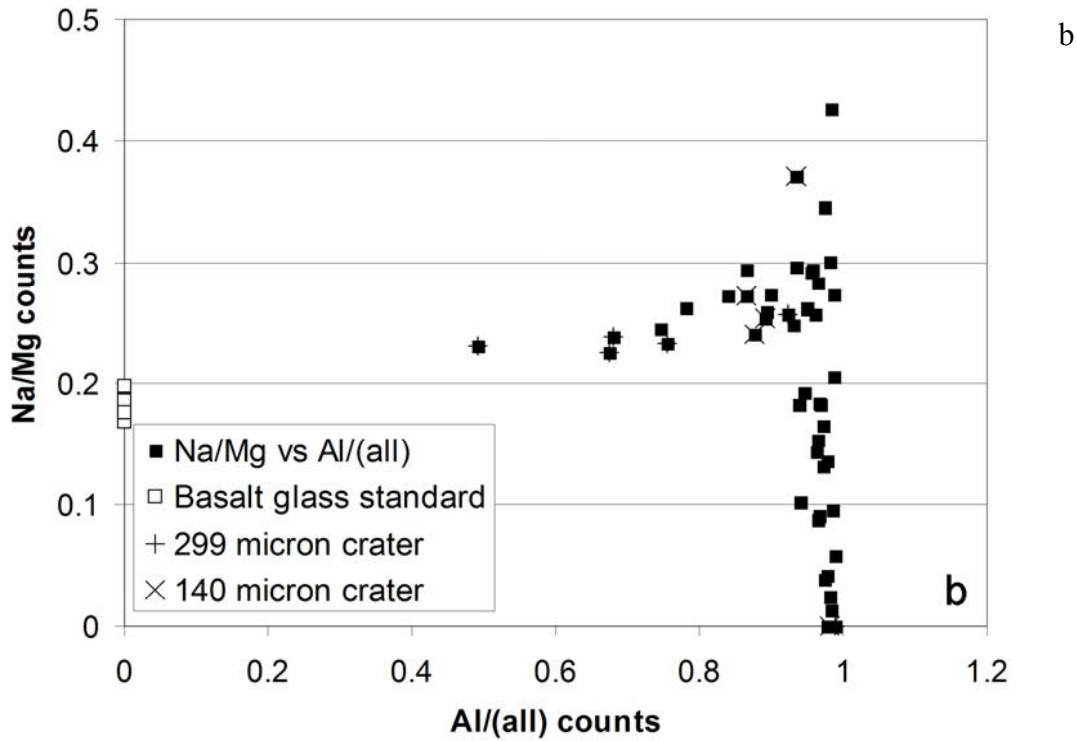

Figure 13. Basalt glass LGG craters, beam to foil perpendicular incidence (beam to residue oblique). EDS X-ray count rates for Na and Mg are plotted against a) crater diameter and b) proportion of Al X-rays in spectrum. Higher Al proportions indicate thinner impact residue, with enhanced signal from metal beneath.

The plot for small basalt glass impact craters (less than 10 microns diameter) in Fig. 13 shows a very wide range of Na/Mg X-ray counts, with many values in excess of that in the original projectile composition, clearly not a true reflection of Na and Mg abundance and probably due to enhanced low energy X-ray escape from a thin inclined residue layer. However, the very low Na/Mg in some very small craters also does suggest that volatile Na may have been lost.

**Elemental fractionation in impact residues**

The high shock pressures and temperatures experienced by particles impacting at high velocity are sufficient to induce changes in structural state, with shock deformation of crystalline structure, melting and even loss as vapour. SIMS studies of spacecraft surfaces returned from LEO show signs of elemental fractionation in the very thin films deposited around features created by hypervelocity impact of micrometeoroids (e.g. Amari et al., 1993). Models such as that of Bernhard et al. (1994b) show that LEO impact velocities (around 20 kms$^{-1}$) can cause peak pressures in excess of 400 GPa, with release temperatures sufficient to vaporise a substantial portion of a silicate or sulfide particle. During impact at 6.1 kms$^{-1}$ (the Stardust encounter relative velocity), the peak pressure may have been much lower, reaching less than 100GPa. This is still sufficient to melt most silicate minerals (Stöffler et al., 1988), although the presence of solid remnants in many of our experimental craters suggests that stress decays rapidly and strain is distributed inhomogeneously within the projectile, due to irregular particle shape and internal structure giving interference from complex free surfaces. However, the potential for elemental fractionation from high temperature melts into vacuum is obvious from the range in evaporation temperature data tabulated by Lodders (2001), with sodium and sulfur largely in the vapour state at temperatures of 900°C. In their pioneering study, Lange et al. (1986) used SIMS to demonstrate strong spatial elemental fractionation in vapour deposits on foils suspended above impact targets. Their results, from experimental shots at velocities similar to the Stardust encounter, highlight the necessity of sampling bulk (micron scale) residue, rather than thin films re-deposited from vapour, if reliable analysis of volatile elements is required.

Using EDS, Hörz et al (1983) reported that a substantial proportion of sodium (Na) (>20%) is lost from impact residues in their LGG impact experiments onto gold, with shock pressures >100GPa. For our simulation of Stardust impacts we chose a basalt glass sample prepared by the United States Geological Survey (USGS). Basalt glass NKT-1G has been used for an international comparison of laboratory analyses (Wilson and Potts 2004; Potts et al., submitted to Geostandards and Geoanalytical Research), has a very well characterised composition and contains volatile Na and potassium (K) at levels adequate for EDS determination. A glass shard was crushed and sieved at NHM to provide projectiles of a wide size range (from 20 to 250 microns). The powder was fired as a buckshot onto Stardust aluminum foil at Canterbury, impacting on Al1100 foil at 6.1 kms$^{-1}$. EDS analyses of residue in situ within crater were performed at NHM to determine whether major elemental fractionation was evident. Impact residue from the lip of a 260 micron diameter crater was extracted and thinned using FIB, and then subsequently analysed by ATEM-EDS at LLNL. The residue composition was compared to a similar ultrathin section cut through a grain of the projectile powder, and analytical results were plotted as elemental ratios relative to calcium (Ca), a refractory element. Ca was chosen rather than Al, so that incorporation of Al alloy might be made obvious, but with Ti as a second high temperature refractory, also expected to show little, if any loss.

Table 5. Comparison of basalt glass projectile and impact crater residue, weight % compositions determined by 12 analyses of each by ATEM on FIB ultrathin sections.

|         | Na   | Mg   | Al   | Si    | P    | K    | Ca   | Ti   | Mn   | Fe   | O     |
|---------|------|------|------|-------|------|------|------|------|------|------|-------|
| Glass   |      |      |      |       |      |      |      |      |      |      |       |
| Average | 2.50 | 7.69 | 4.29 | 15.71 | 0.33 | 0.55 | 4.71 | 0.96 | 0.06 | 3.26 | 59.94 |
| St. dev.| 0.15 | 0.29 | 0.16 | 0.59  | 0.04 | 0.05 | 0.21 | 0.07 | 0.01 | 0.12 | 0.32  |
| Residue |      |      |      |       |      |      |      |      |      |      |       |
| Average | 2.73 | 8.34 | 5.22 | 14.61 | 0.35 | 0.60 | 5.11 | 1.06 | 0.14 | 3.53 | 58.32 |
| St. dev.| 0.22 | 0.38 | 0.30 | 0.63  | 0.07 | 0.10 | 0.36 | 0.13 | 0.22 | 0.29 | 1.61  |

Our earlier, preliminary analyses had suggested that there might be loss of Mg, Si and Fe with each depleted by approximately 10%, hence preserving the important stoichiometric relationship between the major elements of common mafic silicate minerals. Al, Ti, P and K were preserved in the residue at levels very close to the basalt projectiles, but Na was depleted by approximately 28%. However, our second suite of analyses from other parts of the same crater residue (Table 5) now suggest that the only element to be depleted is Si, with Al enhanced (probably due to extremely fine inclusions of the Al alloy melt). A comparison between projectile and residue compositions is shown in Fig. 14.
WDS and EDS analyses of polished sections through the basalt glass show that it is effectively homogeneous at sub-micron scales, and we cannot account for the variation in residue composition across a small crater as a function of projectile heterogeneity. It therefore appears from our results that there may be diverse levels of loss of volatile Na even within a single small crater. Data from craters of greater than 20 microns diameter (Fig. 13) suggest that their thicker residues may show little depletion of Na.

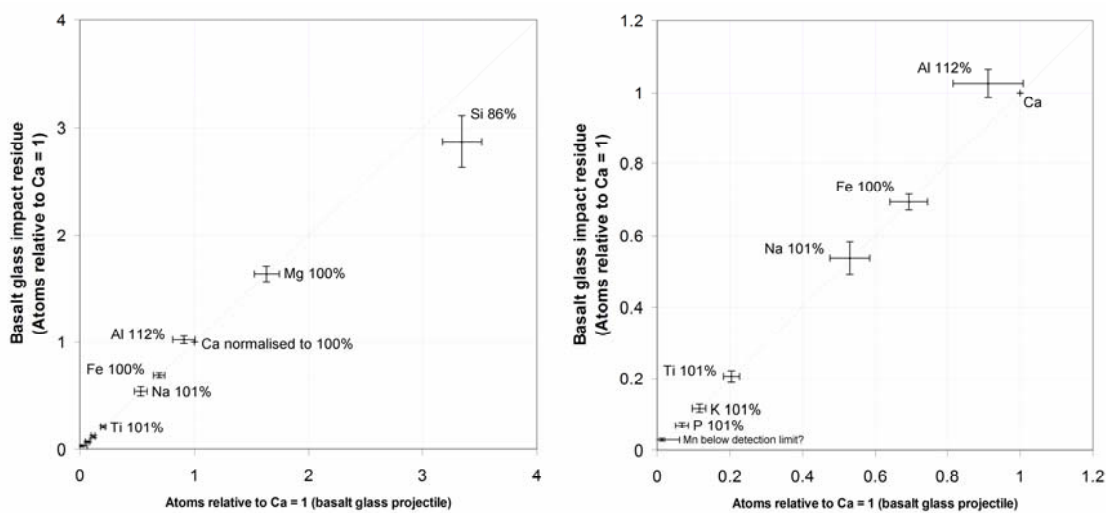

Figure 14. Basalt glass NKT-1, comparison of residue and projectile compositions as determined by ATEM EDS of ultrathin sections cut by FIB at LLNL.

Sulfur is also an important volatile element in interpretation of extraterrestrial materials. We chose to test sulfur loss in impacts of a powder from the NHM pyrrhotite sample BM.2005,M317, shot at 5.97 kms$^{-1}$ onto Al1100 foil. The ratio of SK to Fe L X-ray counts for a suite of craters of 2 to 400 microns diameter is shown in Fig. 15. To distinguish true volatile loss from S Kα absorption (where there was possibility of 'shadowing'), we determined Fe K /Fe L for the small craters. Residues with high Fe K/Fe L are shown as open triangles in Fig. 15. With these data excluded, a clear decline in S Ka with smaller crater size and thinner residue becomes evident, showing volatile S loss is very important for the smaller craters (< 10 microns diameter), but also significant for craters of more than 300 microns diameter.

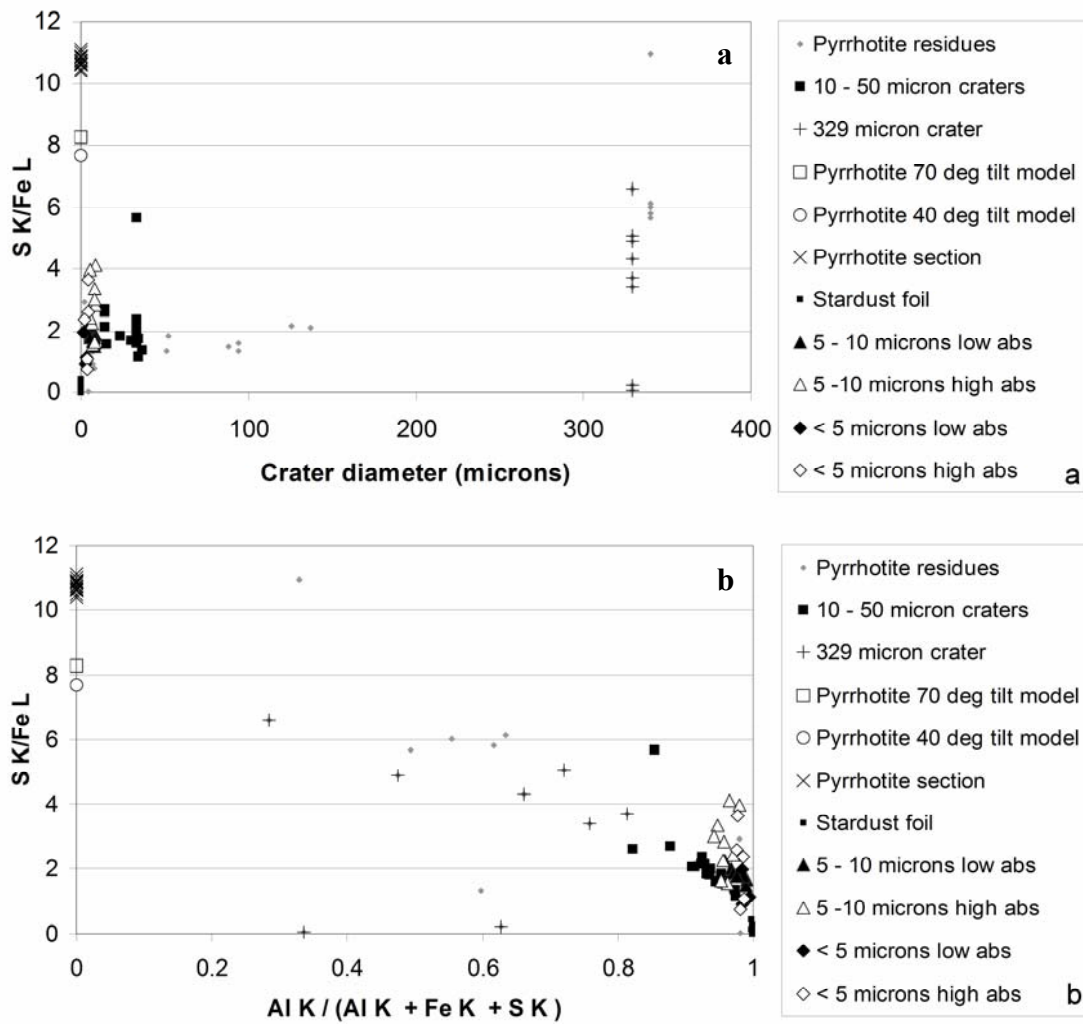

Figure 15. Pyrrhotite LGG impact on Al1100. Plot of S K to Fe L X-ray count ratios as a function of a) crater size and b) residue 'thinness' (proportion of Al X-rays in spectrum). Spectra from Oxford instruments INCA at NHM.

EDS determinations of residue in a tilted crater of 172 microns diameter showed a reduction in sulfur from 39% by weight in the projectile to 33% in the residue (equivalent to loss of 22% of the S atoms relative to Fe). To quantify sulfur loss in smaller craters we used FIB to prepare ultrathin sections of the pyrrhotite projectile and a residue layer within a crater of c. 12 microns diameter (Fig. 16).

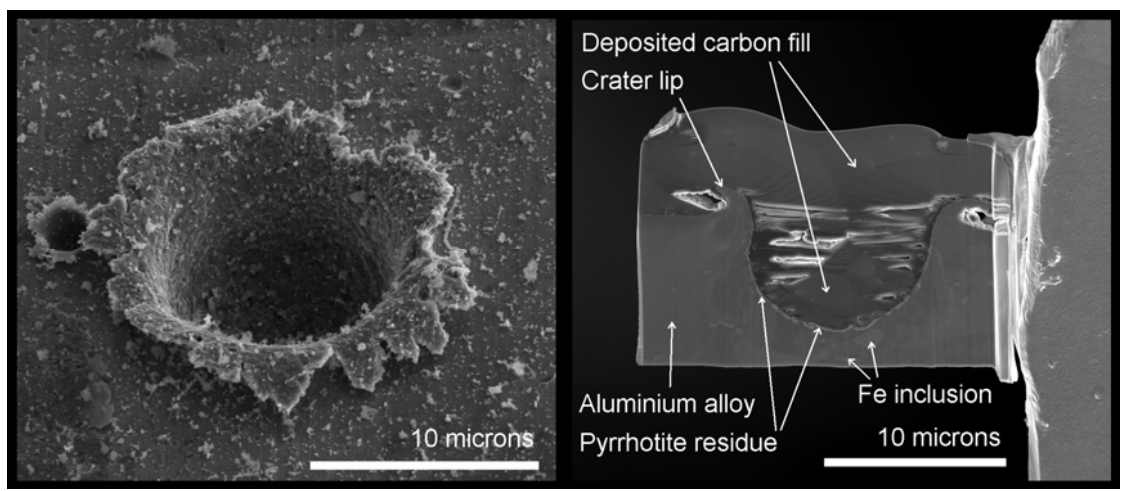

Figure 16. Secondary electron image of crater formed by LGG impact of a small pyrrhotite grain onto Al1100 foil at 5.97 kms$^{-1}$, and extracted ultrathin FIB section.

Table 6. Analyses of pyrrhotite (BM.2005,317) projectiles and impact residue.

| | S wt% | Fe wt% | Fe/S atomic | Ni wt% | Cu wt% | As wt % |
|---|---|---|---|---|---|---|
| Polished section of pyrrhotite chosen for projectiles (8 analyses, Cameca WDS at NHM) | | | | | | |
| Average | 38.65 | 61.38 | 0.91 | 0.02 | 0.02 | 0.11 |
| St. deviation | 0.20 | 0.24 | 0.01 | 0.02 | 0.01 | 0.0 |
| Pyrrhotite projectile, FIB ultrathin section (8 normalised ATEM EDS analyses at LLNL) | | | | | | |
| Average | 38.64 | 61.36 | 0.91 | | | |
| St. deviation | 0.64 | 0.60 | 0.02 | | | |
| Impact residue in LGG crater, FIB ultrathin section (12 normalised ATEM EDS at LLNL) | | | | | | |
| Average | 26.86 | 73.14 | 1.57 | | | |
| St. deviation | 1.75 | 1.75 | 0.15 | | | |

EDS analyses from TEM showed loss of about 30% by weight of the original S (Table 6), an equivalent of loss of 42% of the S atoms relative to Fe. The plot of Fig. 15 suggests that even more S may be lost from craters of smaller than 5 micrometers diameter. Together with the uncertainty as to geometrical effects inherent in very small craters (<10 microns diameter), this may preclude quantitative determination of sulfur by in situ EDS analysis of the smallest intact craters. However, the absorption of sulfur X-rays within a spectrum due to geometrical issues may be recognised if a positive Brehmstrahlung gradient is observed (see above). If a high accelerating voltage is used, a comparison of Fe K and Fe L count rates may also readily reveal absorption effects.

**Conclusions**

Stardust foil craters can be interpreted in terms of impacting particle size, density and, in many cases, composition. Impact by solid particles of low porosity can be recognised by their relatively simple bowl morphology, although the detailed internal shape may reflect irregularity in the particle shape. For a bowl-shaped crater, depth to diameter ratio may be used to estimate particle density, and together with compositional information from residue, to decide upon an appropriate crater diameter to particle diameter calibration. Although it is not yet possible to generate laboratory impacts by such projectiles, it is likely that grains with an aggregate structure containing discrete mass concentrations (sub-grains), will not create a single bowl shaped crater, but instead irregular patches formed by overlapping fields of interfering bowl shapes.

Energy dispersive X-ray microanalysis will yield a useful inventory of elements present in most craters, but it is important to be aware of potential contamination by inclusions in the foil (mainly Fe, but also including Si, Ti, V, Ni and Cu). For impacts onto Al1100 foils, in situ quantitative analysis of residues within larger craters (>10 microns in diameter) can yield very good, reliable element ratios, but it is still important to avoid geometrical problems with X-ray absorption, which can be achieved by tilting the substrate to present perpendicular incidence of the electron beam on the residue. However, accurate ratios of elements cannot be determined reliably from areas shown to be affected by 'shadowing'. In situ SEM-EDS quantitative analysis of residues preserved in craters of less than 10 microns diameter becomes much more difficult, due to complex absorption phenomena that are specific to the location of each single spectrum and cannot be fully evaluated. For small craters, analytical transmission electron microscopy of ultra-thin sections prepared by FIB is required to produce good determinations of element ratios.

When interpreting cometary chemical composition based upon analyses of residues formed by impact onto Al1100 foil at c. 6 kms$^{-1}$, it is important to consider loss of volatile elements such as sodium and sulfur. Our experiments have shown a loss of about 25% by weight of sulfur from an iron sulfide projectile impacted under these conditions. The plotted data in this paper could be applied as an approximate correction for sulfur loss from residues in craters larger than 10 microns in diameter. In smaller craters with thinner residue layers, volatile loss for both sulfur and sodium is most severe, and is probably so variable as to prevent use of a reliable correction factor.

**Acknowledgements:**
We wish to thank Fred Hörz at NASA/JSC for his advice and support in our experiments to provide calibration data for Stardust and for the provision of foil target material. Mark Burchell and Mike Cole gratefully acknowledge PPARC research funding to provide light gas gun facilities. Anton Kearsley, John Spratt and Gary Jones thank NHM for provision of electron microscope, microprobe and ICP facilities. Tony Wighton (NHM) prepared excellent polished sections for analysis and mapping of the foils and projectile materials. Staff at LLNL acknowledge that their work was performed in part under the auspices of the US Department of Energy, National Security Administration by the University of California, Lawrence Livermore National Laboratory under contract No. W-7405-Eng-48.